\documentclass[aps,prl,twocolumn,showpacs,superscriptaddress,groupedaddress]{revtex4}  
\usepackage{graphicx}  
\usepackage{epsfig}
\usepackage{dcolumn}   
\usepackage{bm}        
\usepackage{amssymb}   
\usepackage{amsmath}
\usepackage{hyperref}
\usepackage{epstopdf}
\usepackage{color}
\usepackage{cancel}
\begin{document}

\title{Optimal linear optical implementation of a single-qubit damping channel}

\date{\today} 

\author{Kent Fisher}\email{k8fisher@uwaterloo.ca}
\affiliation{Institute for Quantum Computing, Department of Physics and Astronomy, University of Waterloo, Waterloo, N2L 3G1, ON, Canada}

\author{Robert Prevedel}
\affiliation{Institute for Quantum Computing, Department of Physics and Astronomy, University of Waterloo, Waterloo, N2L 3G1, ON, Canada}

\author{Rainer Kaltenbaek} \affiliation{Institute for Quantum Computing, Department of Physics and Astronomy, University of Waterloo, Waterloo, N2L 3G1, ON, Canada}
\affiliation{Vienna Center for Quantum Science and Technology, Faculty of Physics, University of Vienna, Boltzmanngasse 5, 1090 Vienna, Austria}

\author{Kevin J. Resch}
\affiliation{Institute for Quantum Computing, Department of Physics and Astronomy, University of Waterloo, Waterloo, N2L 3G1, ON, Canada}

\begin{abstract}
We experimentally demonstrate a single-qubit decohering quantum channel using linear optics. We implement the channel, whose special cases include both the amplitude-damping channel and the bit-flip channel, using a single, static optical setup. Following a recent theoretical result [M. Piani \emph{et al.}, Phys. Rev. A, \textbf{84}, 032304 (2011)], we realize the channel in an optimal way, maximizing the probability of success, i.e., the probability for the photonic qubit to remain in its encoding. Using a two-photon entangled resource, we characterize the channel using ancilla-assisted process tomography and find average process fidelities of $0.9808 \pm 0.0002$ and $0.9762 \pm 0.0002$ for amplitude-damping and the bit-flip case, respectively.
\end{abstract}

\maketitle

\emph{Introduction.} Time evolution in quantum mechanics converts a density matrix to another density matrix. This evolution is referred to as a quantum channel and can be described mathematically as a completely positive (CP) map \cite{Nielsen2000}. Because of the generality of the concept of quantum channels, their use is ubiquitous in quantum information. For example, unitary quantum channels are used in quantum computing to describe quantum gates. Non-unitary channels, on the other hand, describe the interaction of quantum states with an environment, and have recently been connected to fundamental physical questions in quantum information science, such as channel capacity, superadditivity \cite{Smith2008, Hastings2009} and bound entanglement \cite{Entanglement1998}.

Linear optics and single photons have several characteristics that make them an ideal testbed for quantum information. Single-qubit unitaries are easy to implement as, for polarization encoded qubits, they only require waveplates. Photonic qubits also exhibit long coherence times, and spontaneous parametric down conversion allows the generation of high-quality entangled states, which can be easily manipulated. However, certain operations, such as the two-qubit CNOT-gate, are difficult in this architecture \cite{Milburn1989,Lutkenhaus1999}, and can only be implemented probabilistically~\cite{Knill2001, Ralph2002, OBrien2003}.

Unfortunately, the ease of single-qubit operations does not extend to more general CP maps. Some quantum channels, like the depolarizing single-qubit channel \cite{Nielsen2000} can be implemented with unit probability, but this is not the case in general. For instance, the amplitude-damping channel, a non-unital quantum process, has been implemented in linear optics only with a limited success probability of $1/2$ \cite{Qing2007,Lee2011}. In one experiment~\cite{Almeida2007}, the interaction of a qubit with ancilla modes was implemented such that the resulting counts replicated those expected for an amplitude-damping channel. However, the quantum information did not remain in its original encoding and cannot be viewed as a single-qubit channel. Recently, it was shown by Piani \emph{et al.} \cite{Piani2010} that, \emph{any} single-qubit quantum channel could be implemented probabilistically using linear optics and postselection, i.e., similar to many two-qubit operations. Moreover, they proved that such implementations can be achieved with the optimal success probability.

In the present work, we use this recent theoretical result to design and demonstrate a linear-optics-based implementation of a certain class of non-unital single-qubit quantum channels called ``damping channels". The class of channels we focus on can be parametrized by two real numbers: $\alpha, \beta$. In the operator-sum representation, the channel's action on an arbitrary quantum state $\rho$ can be written as $\mathcal{E}(\rho) = \sum_i A_i \rho A^{\dagger}_i$, where the two Kraus operators are \cite{Wolf2007}:
\begin{equation}
\label{equ:kraus}
A_0 = \left ( \begin{array}{cc}
\cos{\alpha} & 0 \\
0 & \cos{\beta}
\end{array} \right )
,
A_1 = \left ( \begin{array}{cc}
0 & \sin{\beta} \\
\sin{\alpha} & 0
\end{array} \right )
\end{equation}

\noindent This channel is of great interest as its special cases include the amplitude-damping ($\alpha = 0)$ and bit-flip ($\alpha = \beta$) channels, both of which are common sources of error in other implementations of quantum information processing, such as ion traps. Furthermore, it belongs to the small class of quantum channels for which the quantum capacity can be directly calculated via the coherent information~\cite{Devetak2005}.

\begin{figure}[t!]
\centering
\epsfig{file=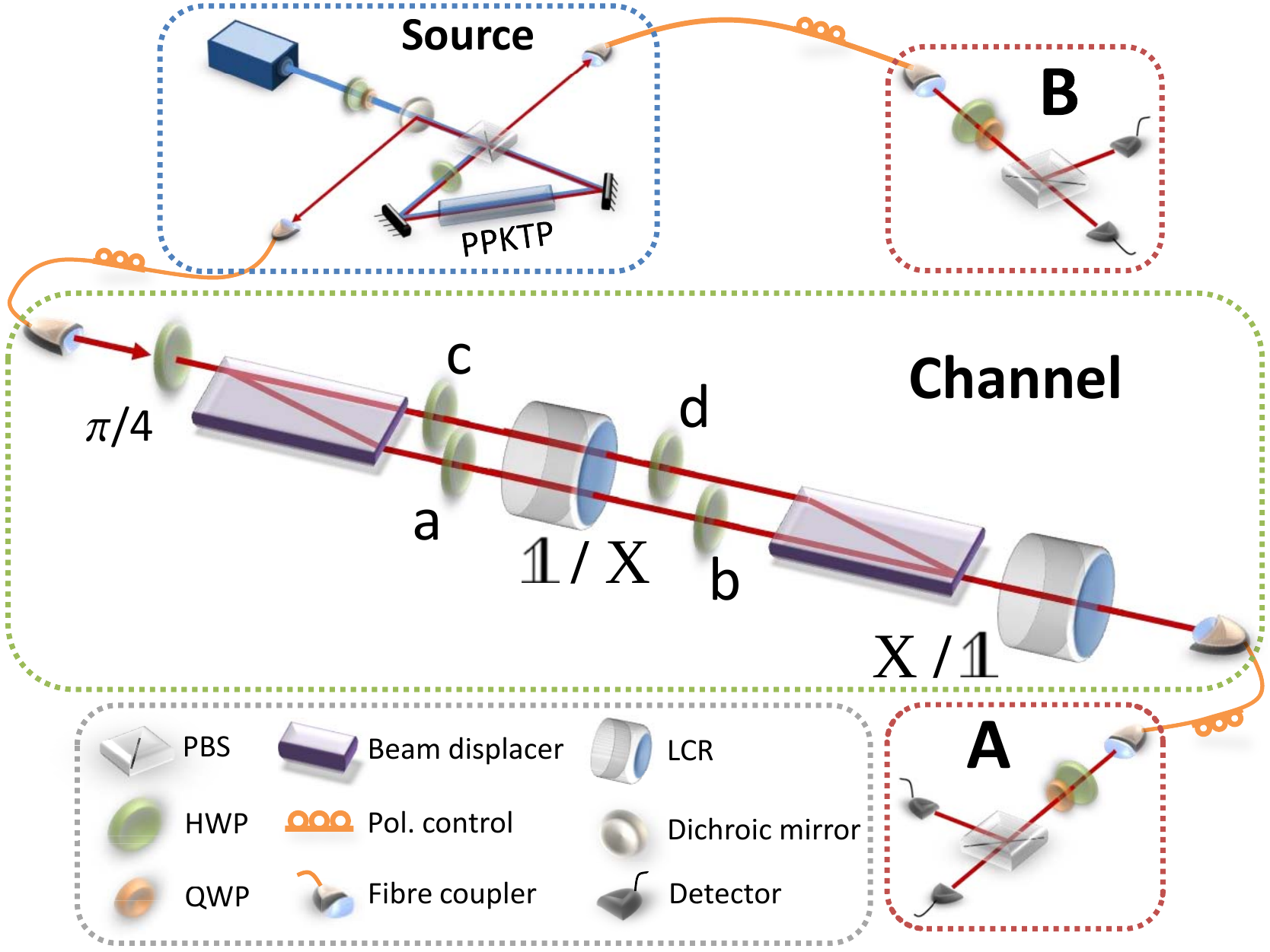,width=0.85\linewidth}
\caption{The experimental setup. We use spontaneous parametric down conversion in periodically poled KTiOPO$_4$ (PPKTP) to generate entangled photon pairs of the form $\left | \Phi^{+} \right \rangle = \frac{1}{\sqrt{2}} \left(\left|H H\right\rangle + \left|V V\right\rangle\right)$ which are subsequently coupled into single-mode fibres. One of the photons is sent through the damping channel parameterized by $\alpha$ and $\beta$, which are set by the angles $a$, $b$, $c$ and $d$ of four half-wave plates (HWPs). Two liquid-crystal retarders (LCRs) switch anti-correlatively between the identity, $\openone$, and the Pauli $X$ operation. The polarization of each photon is measured by an analyzer (A and B) consisting of a half- and a quarter-wave plate (QWPs) followed by a polarizing beam splitter (PBS). Eventually, the photons are detected by single-photon counting modules.}\label{fig:setup}
\end{figure}

Here, we experimentally realize this single-qubit damping quantum channel using linear optics. We can use the setup to add controlled amounts of noise of various types to a single qubit. The schematics of the experimental setup are shown in Fig.~\ref{fig:setup}. The key step in the implementation of the channel is the splitting of the polarization encoded information into different spatial modes, which then allows for the manipulation of different logical states independently of one another. An arrangement of half-wave plates and liquid-crystal retarders allows us to probabilistically implement both Kraus operators within a single, static optical setup. We characterize the channel using a new ancilla-assisted quantum process tomography method, and show the optimality of our optical implementation, with our success rates in the amplitude-damping case surpassing those of previous implementations \cite{Qing2007, Lee2011}. In order to characterize the action of the channel on entanglement we study the amount of entanglement of photon pairs when one photon is sent through the channel.

\emph{Optimality of the implementation.}
Following Ref.~\cite{Piani2010}, it can be shown that the probability of success for a specific Kraus decomposition $\{A_i\}$ is $p_{\text{succ}}(\{A_i\}) = \left(\sum_i \left\| A_i \right\|^2_\infty\right)^{-1}$, where the norm $\left\| M \right\|_\infty$ is the largest singular value of the operator $M$. Maximizing over all possible Kraus decompositions ${A_i}$ describing the channel allows to achieve the optimal success probability $p_{\text{succ}}^{\text{opt}} = \max_{A_i} \frac{1}{\sum_i \left\| A_i \right\|^2_\infty}$. For our particular channel, if we assume that $\cos(\alpha)\geq\cos(\beta)$, this yields:
\begin{equation}\label{equ:psucc}
    p_{\text{succ}}^{\text{opt}} = \frac{1}{\cos^2{\alpha} + \sin^2{\beta}  }
\end{equation}
In order to achieve $p_{\text{succ}}^{\text{opt}}$, we have to implement each Kraus operator with individual probabilities $p_{A_i}= \left \| A_i \right \|^2_{\infty}\cdot p_{\text{succ}}^{\text{opt}}$. We find that the optimal probability of success is achieved for $p_{A_0} = \frac{\cos^2{\alpha}}{\cos^2{\alpha} + \sin^2{\beta}}$ and $p_{A_1} = \frac{\sin^2{\beta}}{\cos^2{\alpha} + \sin^2{\beta}}$. We show experimentally that for various values of $\alpha$ and $\beta$, which can be independently controlled in our experiment, we indeed achieve this upper bound.

\emph{Ancilla-assisted quantum process tomography.}
Quantum process tomography (QPT) allows to experimentally reconstruct the superoperator describing an unknown physical process. Ancilla-assisted QPT (AAQPT) uses ancillary qubits to facilitate the reconstruction procedure for quantum state measurements. It has also been shown \cite{Altepeter2003} that AAQPT gives decreasing statistical errors as the amount of entanglement between the primary and ancilla systems is increased.

AAQPT has been used to study various unitary quantum gates \cite{Altepeter2003} but has not yet been extended to the characterization of non-unital channels. In our work, and in contrast to previous AAQPT schemes \cite{Chow2009, James2001, OBrien2004}, we do not assume an ideal description of our initial state, but rather measure and include it when using a maximum-likelihood technique to find the physical matrix that best describes the action of the experimentally implemented channel.

The standard techniques for QPT and AAQPT are described in \cite{Nielsen2000} and \cite{Altepeter2003}, respectively. Below, we outline our method following their nomenclature. Consider a two-qubit state, $\rho_{AB}$, whose density matrix is known; e.g., it might have been reconstructed using quantum state tomography (QST). The quantum channel $\mathcal{E}$ acts on subsystem A, while subsystem B is unaffected. The transformed two-qubit state after the channel is $\rho^\prime_{AB} = \left ( \mathcal{E} \otimes \openone \right ) (\rho_{AB})$. Characterizing $\rho^\prime_{AB}$, e.g., by performing standard QST, allows for  reconstruction of the process using the Choi--Jamio\l kowski isomorphism \cite{Choi1975, Jamiolkowski1972}.

The quantum process can then be written as $\rho^\prime_{AB} = \sum_{m,n=0}^{d^2-1} \chi_{mn} ( \tilde E_m \otimes \openone ) \rho_{AB} ( \tilde E_n \otimes \openone ) ^{\dagger}$ where $\{ \tilde E_i \}$ are operators which form a basis in the space of $d \times d$ matrices ($d=2$ in our case). It is common to use the basis formed by the Pauli matrices $\left \{ \openone, X, Y, Z \right \}$. The $d^2$-dimensional process matrix $\chi$ then fully describes the quantum process. In our maximum-likelihood technique, we parameterize $\chi$ by $d^4 - d^2 = 16$ real numbers \cite{OBrien2004, James2001} and seek to minimize the following function:
\begin{align}
\label{equ:ml}
f = &\sum^{\nu}_{i=1}  \frac{ ( n_{i} - \mathcal{N} \text{Tr} [ M_i \rho^\prime_{AB} ] )^2 }{2 \mathcal{N}  \text{Tr}  [ M_i \rho^\prime_{AB} ]} + \nonumber \\
    &\lambda  \sum_k \left [ \sum_{m,n} \chi_{mn} \text{Tr}( \tilde{E}^{\dagger}_n \tilde{E}_m \tilde{E}_k  ) -  \text{Tr}(\tilde{E}_k)   \right ]^2,
\end{align}

\noindent such that the resulting $\chi$ most closely resembles a physical quantum process. Here, $i$ labels the measurement setting in the final QST, $\nu$ is the number measurement settings, $n_{i}$ is the number of two-fold coincidence counts recorded in the $i^{\text{th}}$ setting, $\mathcal{N}$ corresponds to the number of photons incident on the detectors, $M_i$ is the projector of the $i^{\text{th}}$ measurement, and $\lambda$ is a Lagrange multiplier used to force the resulting process matrix to be trace preserving \cite{Chow2009}.

\begin{figure}[t!]
\centering
\epsfig{file=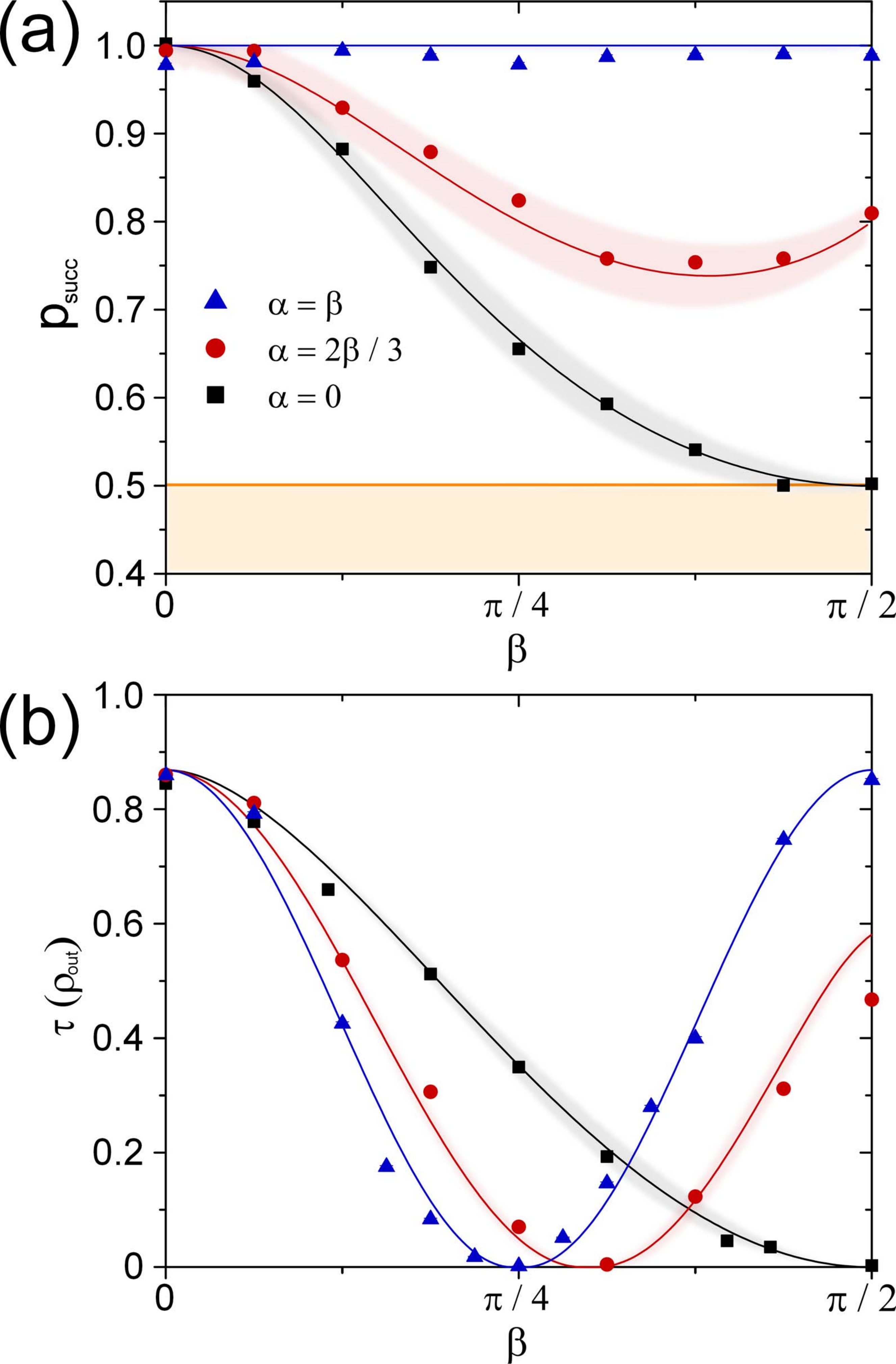,width=0.85\linewidth}
\caption{ (a) Probability of success as a function of damping parameter $\beta$ for cases $\alpha = 0$, $\alpha = \frac{2}{3} \beta$ and $\alpha = \beta$. The shaded region below 0.5 represents probabilities of success in previous optical implementations of the amplitude-damping channel, see Refs.~\cite{Qing2007,Almeida2007,Lee2011} (b) The tangle, $\tau$, of the resulting two-photon state as a function of damping $\beta$ after one photon has passed through the damping channel. Errors in the experimental data are calculated from Poissonian noise in the coincidence counts and are not visible on the scale of the plots. The solid lines in both panels represent the theoretically expected dependance. In (b), the experimental density matrix of the input state was used for the calculations. The shaded regions around the theory curves in both plots represent the expected standard deviation assuming $1^{\circ}$ and $1\%$ rotation errors in the HWPs and LCRs respectively. The margin of error in the $\alpha = \beta$ case is significantly smaller than for the other cases due to the fact that the HWP angles and LCR settings all lie at points where partial derivatives of the Kraus-operator matrix elements are zero.}\label{fig:psucctangle}
\end{figure}

\emph{Experiment.}
We use the experimental setup shown in Fig.~\ref{fig:setup} to implement the quantum channel defined by the Kraus operators in Eq.~\ref{equ:kraus}. Two $40\,$mm calcite beam displacers are used to construct an interferometer. Within these beam displacers, photons with  horizontal ($\left | H \right \rangle$) and vertical ($\left | V \right \rangle$) polarization are spatially displaced with respect to each other \cite{OBrien2003}. Half-wave plates (HWPs) are used to set the amount of damping by allowing to adjust $\alpha$ and $\beta$ in Eq.~\ref{equ:kraus}. The relations between these parameters and the individual angles $a,b,c,d$ of the four HWPs are given by $\sin{4a} = \frac{\cos{\beta}}{\cos{\alpha}}$, $b = a -\frac{\pi}{4}$, $\sin{4c} = -\frac{\sin{\alpha}}{\sin{\beta}}$ and $d = \frac{\pi}{2} - c$. The channel is realized by switching randomly between the Kraus operators $A_0$ and $A_1$. The switching is performed using two liquid-crystal retarders (LCRs). We set the two LCRs to $X_1$ and $\openone_2$, respectively, to implement $A_0$, and we set them to $\openone_1$ and $X_2$ to implement $A_1$. Here, the subscripts represent the action of the first and the second LCR. The probabilities, $p_{A_0}$ and $p_{A_1}$, with which each configuration is realized are determined by the values of $\alpha$ and $\beta$ such that the overall success probability of realizing the channel is optimal \cite{Piani2010}. The switching rate of the LCRs was chosen to be $10\,$Hz, significantly faster than the integration time for a single measurement ($5\,$s).

To characterize the channel, we use the AAQPT scheme outlined above. Our resource state is an entangled photon pair generated in a type-II down conversion source in a Sagnac configuration~\cite{Fedrizzi2007, Prevedel2011}. A 0.5\,mW laser diode at 404.5\,nm pumps a 25 mm periodically-poled crystal of KTiOPO$_4$ (PPKTP). This typically yielded a coincidence rate of 10\,kHz. The characterization of the channel is executed as follows: The HWP angles $a$, $b$, $c$ and $d$ are set to zero and the LCRs to $X_1$ and $\openone_2$ such that the channel acts as the identity map. A QST is performed to obtain the density matrix of the input state, $\rho_{AB}$. The HWP angles and the probabilities for switching the LCRs and the HWP angles are then set according to the values of $\alpha$ and $\beta$. Another QST yields the output state, $\rho^\prime_{AB}$. QST involves recording coincidences for all combinations of the eigenstates of the Pauli X, Y and Z operators. For each of these 36 projective measurements, we integrated coincidence counts for 5\,s. The resulting data were then used in conjunction with Eq.~\ref{equ:ml} to reconstruct the superoperator describing the quantum process.

\emph{Results.}
We now turn to our main result, the optimality of our quantum channel implementation. Fig.~\ref{fig:psucctangle}a shows the probability of success for the amplitude-damping, bit-flip, and one in-between case ($\alpha = \frac{2}{3} \beta$). Since amplitude-damping manifests itself as photon loss in our particular implementation, determining the probability of success reduces to measuring the transmission of the channel. The experimental data closely follow the theoretical predictions (solid lines) that are based on Eq.~\ref{equ:psucc}.

Previous optical implementations of the amplitude-damping channel \cite{Qing2007,Lee2011} have given at most 50\% probability of success~\cite{footnote}, whereas here we find that only in the case of maximum damping ($\beta = \pi/2$) the probability of success decreases to 50\%. The experimental results for the success probability closely resemble the theoretical prediction. This is also true for our experimental implementation of the bit-flip channel and the $\alpha = \frac{2}{3}\beta$ case of single-qubit damping.

\begin{figure}

\centering
\epsfig{file=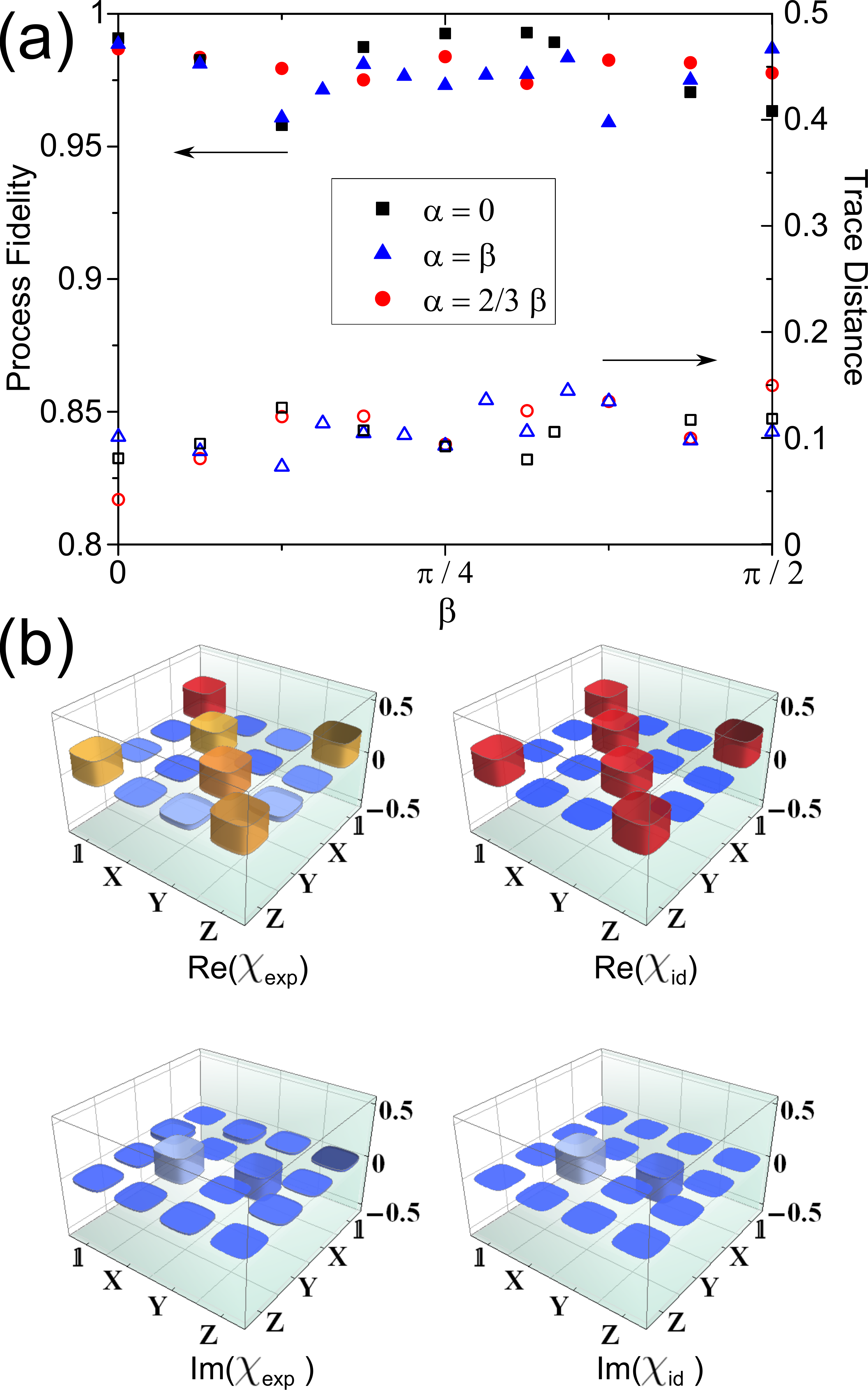,width=0.85\linewidth}
\caption{ (a) Measured process fidelity (solid data points) and trace distance (unfilled data points) as a function of damping for each of the three cases studied. Error bars ($\sim 10^{-3}$), calculated using Monte-Carlo simulations adding Poissonian noise to the measured state tomography counts in each run, are too small to see on this scale. (b) Real and imaginary parts of the experimentally determined and ideal process matrices at maximum amplitude-damping ($\alpha = 0, \beta = \pi/2$).} \label{fig:trd}
\end{figure}

Fig.~\ref{fig:psucctangle}b shows the tangle~\cite{Wooters1998} of the two-photon output density matrix, $\rho'_{AB}$, for the amplitude-damping, bit-flip, and $\alpha = \frac{2}{3} \beta$ cases, where one of the two photons passes through the quantum channel. Theoretical curves are based on the action of the respective ideal quantum channels on the experimental input state. It can be seen in each case, that the experimental data agrees well with the theoretical prediction, showing that the experimentally implemented channel closely resembles the ideal one. The process fidelity is defined by $\mathcal{F} = \text{Tr} \sqrt{ \sqrt{\chi_{\text{exp}}} \chi_{\text{id}} \sqrt{\chi_{\text{exp}}} }$~\cite{Jozsa1994}, where $\chi_{\text{exp}}$ and $\chi_{\text{id}}$ are the experimental and ideal process matrices, respectively. We find process fidelities of $0.9808 \pm 0.0002$, $0.9762 \pm 0.0002$, $0.9805 \pm 0.0003$ for the $\alpha = 0$, $\alpha = \beta$ and $\alpha = \frac{2}{3} \beta$ cases of the channel, respectively. We also compute the so-called maximum trace distance~\cite{Nielsen2000}, which is defined as $\mathcal{D}=\max_{\rho_{\text{in}}} \frac{1}{2} \text{Tr} \left | \rho^{\text{exp}}_{\text{out}} - \rho^{\text{id}}_{\text{out}} \right |$, where $\left | A \right | = \sqrt {  A^{\dagger} A }$ and $\rho^{\text{exp/id}}_{\text{out}} = \sum_{m,n} \chi^{\text{exp/id}}_{mn} \tilde{E}_m \rho_{\text{in}} \tilde{E}^{\dagger}_n$. Operationally, $\mathcal{D}$ corresponds to the highest probability of distinguishing between the experimental and ideal channels using the best possible input state. The average maximum trace distance over all damping values measured were found to be $0.1028 \pm 0.0009$, $0.1077 \pm 0.0006$, $0.1075 \pm 0.0009$ for the three respective cases. In Fig.~\ref{fig:trd} we show both the process fidelity and maximum trace distance as a function of damping for each case, as well as the reconstructed $\chi$ matrix for $\alpha = 0$, $\beta = \pi/ 2$.

\emph{Summary.}
Decoherence plays an important role in quantum information science. Investigating its effects requires careful and well-controlled implementations of these noisy processes. Non-unital damping channels, like the ones studied here, are crucial in further understanding quantum communication, in determining channel capacities and for the generation of bound-entangled states. We have implemented a general damping single-qubit quantum channel with linear optics in which both type and amount of decohering noise can be precisely controlled. A single, static optical setup can perform as the amplitude-damping channel, the bit-flip channel, or more general cases characterized by two real parameters, $\alpha$ and $\beta$. Most importantly, we have shown that this channel has been implemented in an optimal way, so as to maximize the probability of success. The channels were characterized using a new approach to ancilla-assisted process tomography and, in all cases, operate with high fidelity.

\emph{Acknowledgements.} We acknowledge fruitful discussions with J. Lavoie, M. Piani and N. L\"{u}tkenhaus, and are grateful for financial support from Ontario Ministry of Research and Innovation ERA, QuantumWorks, NSERC, OCE, Industry Canada and CFI. R.P. acknowledges support by the Ontario MRI and the Austrian Science Fund (FWF).

\bibliographystyle{apsrmp}

\end{document}